\documentclass[10pt,letterpaper]{iopart}
\usepackage{fullpage}
\usepackage{graphicx}
\usepackage{mathpazo}
\usepackage{cite}

\expandafter\let\csname equation*\endcsname\relax
\expandafter\let\csname endequation*\endcsname\relax\usepackage{amsmath}
\usepackage{amsfonts}
\usepackage{xfrac}

\newcommand{\ssep}{}

\begin{document}
\title{Lattice gases with a point source}
\author{P. L. Krapivsky$^{1}$ and Darko Stefanovic$^{2,3}$}
\address{$^{1}$Department of Physics, Boston University, Boston, Massachusetts 02215, USA}
\address{$^{2}$Department of Computer Science, University of New Mexico, MSC01 1130, 
1 University of New Mexico, Albuquerque, NM 87131-0001, USA}
\address{$^{3}$Center for Biomedical Engineering, University of New Mexico, MSC01 1141, 
1 University of New Mexico, Albuquerque, NM 87131-0001, USA}
\eads{\mailto{paulk@bu.edu},\mailto{darko@cs.unm.edu}}

\begin{abstract}
We study diffusive lattice gases with local injection of particles,
namely we assume that whenever the origin becomes empty, a new
particle is immediately injected into the origin. We consider two
lattice gases: a symmetric simple exclusion process and random
walkers. The interplay between the injection events and the positions
of the particles already present implies an effective collective
interaction even for the ostensibly non-interacting random walkers. We
determine the average total number of particles entering into the
initially empty system. We also compute the average total number of
distinct sites visited by all particles, and discuss the shape of the
visited domain and the statistics of visits.
\end{abstract}

\pacs{02.50.-r,  66.10.C-, 05.70.Ln}

\section{Introduction}
\begin{figure}
\begin{center}
\includegraphics[height=0.4\textwidth]{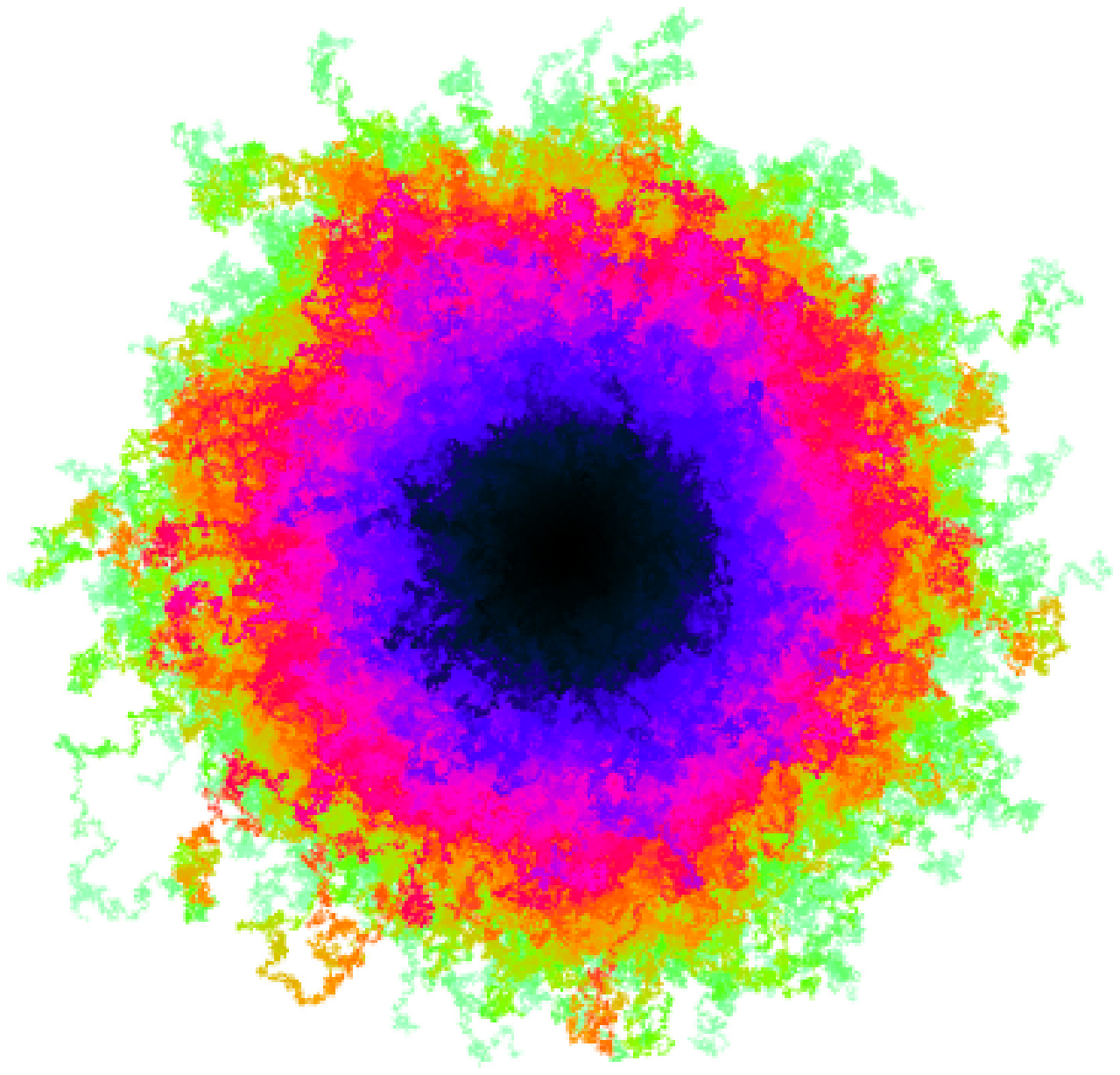}\\
(a)\\
\includegraphics[height=0.4\textwidth]{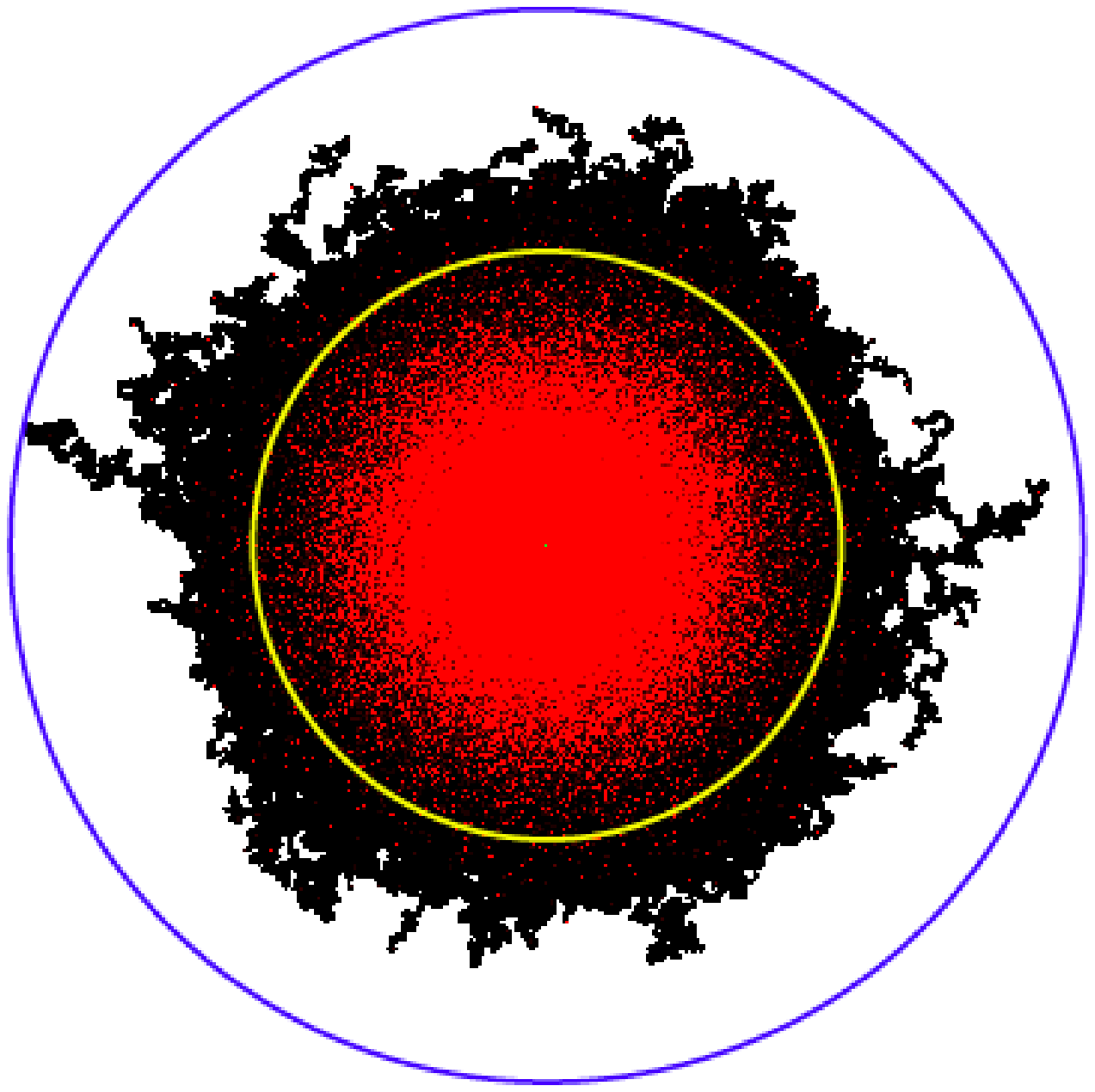}\\
(b)\\
\caption{ [images downsampled for arXiv] 
  (a) Snapshot of a single realization of the two-dimensional SEP at time $t=10^5$. Colour or grey scale indicates first visit time, from dark blue (early) to light green (recent).
  (b) A different view of the same SEP at time $t=10^6$: visited sites in black; current positions of the particles shown as red dots. In yellow, centred at the origin, the largest inscribed disc (no unvisited sites inside); in blue, the smallest circumscribed disc (no visited sites outside). 
\label{fig:domain-example}
}
  \end{center}
\end{figure}

Random walks on a lattice are a common abstraction for particle
diffusion~\cite{Berg1993,Weiss1994,Kolomeisky:2007}.  The
symmetric exclusion process (SEP) is a random walk of multiple
particles subject to exclusion: two particles cannot simultaneously
occupy the same site~\cite{Krapivsky2010book}. Many natural processes,
such as foraging~\cite{PhysForaging}, involve point sources of
particles; the SEP with a localized source corresponds to
monomer-monomer catalysis and the voter
model~\cite{Krapivsky1992,Frachebourg1996,Mobilia2003}.  We study the
SEP with an infinitely strong, unbounded source: whenever the origin
is clear, a new particle is immediately injected.  We also study the
random walk process (RW), in which multiple particles are not subject to exclusion, 
with analogous injection into the origin 
only whenever it is empty.  Using a
combination of analytical results and extensive numerical simulations,
here we derive asymptotic expressions for the number of particles in
the system for all dimensions $d$. We derive expressions for the
number of distinct visited sites and the total visit activity, which
are of interest in models of foraging and
spreading~\cite{Larralde1992a,Larralde1992b,Sastry1996,Yuste1999,Yuste2000}.
In many cases these quantities converge to their asymptotic behaviour
exceedingly slowly, especially in experimentally relevant dimensions
$d=2,3$, therefore simulations should be used to predict them in
short-time applications.  For $d=2$ the domain of visited sites,
Figure~\ref{fig:domain-example}, exhibits a fractal-like boundary
within an annulus surrounding a compact disc; the relative thickness
of the annulus decays logarithmically.

We begin in Section~\ref{sec:SEP-announce} with a detailed
description of the symmetric exclusion process with a point source (SEP),
recapitulating previous results~\cite{Krapivsky2012} (the number of
particles injected) and stating our
new results (the number of visited sites and the statistics of visits) 
along with informal arguments. In Section~\ref{sec:RW-announce}
we introduce the exclusion-free analogue (RW). More thorough
derivations are given in Section~\ref{sec:V-derivations} for the number of
visited sites, and in Section~\ref{sec:A-derivations} for the statistics of visits.

\section{Symmetric exclusion process (SEP) with a point source}
\label{sec:SEP-announce}

A lattice gas of identical particles undergoing nearest-neighbour
symmetric hopping with the constraint that each lattice site can be
occupied by at most one particle at a time is a paradigmatic
interacting particle system, known as the symmetric exclusion
process (SEP). Despite its simplicity, the SEP, along with its
asymmetric cousin the ASEP (in which the hopping rates differ
depending on the direction), exhibits a surprisingly rich set of
dynamical behaviours. These lattice gases have been extensively
investigated~\cite{Spohn1991,KipnisLandim1999,Liggett1999,Schuetz2000,Blythe2007,Derrida2007,Krapivsky2010book},
particularly in one dimension, the most tractable setting. The SEP and
ASEP have been applied to low-dimensional transport of particles with
hard-core interactions, e.g., to the motion of molecular motors along
cytoskeletal fibres~\cite{Chou2011}.  The spreading of very thin
wetting films has been described by a SEP-like
model~\cite{Popescu2012}; here the natural dimensionality of the
substrate is $d=2$ and the average injected mass has been computed for
$d\leq 2$ and shown to compare favourably with experimental
observations~\cite{Burlatsky1996a,Burlatsky1996b}.  We arrived at the
present problem from the question of target search~\cite{Semenov2013} by a stream of
catalytic DNA walkers~\cite{Pei2006,Lund2010,Krapivsky:2007b,Semenov:2011,Olah2013} released
onto a substrate.

Here we analyse the SEP with a localized source on the $d$-dimensional
hypercubic lattice $\mathbb{Z}^d$. We set the total hopping rate to
unity, so the hopping rate to each of the $2d$ nearest-neighbour
sites is equal to $\sfrac{1}{2d}$. A hopping event is allowed only when
the chosen destination is empty.  For the localized source, particles
are injected into the origin.  The source is infinitely strong --
whenever a particle at the origin hops to a neighbouring empty site,
the origin is instantaneously reoccupied by a fresh particle.

\subsection{The SEP model}
\begin{figure}
\begin{center}
\includegraphics[width=0.5\textwidth]{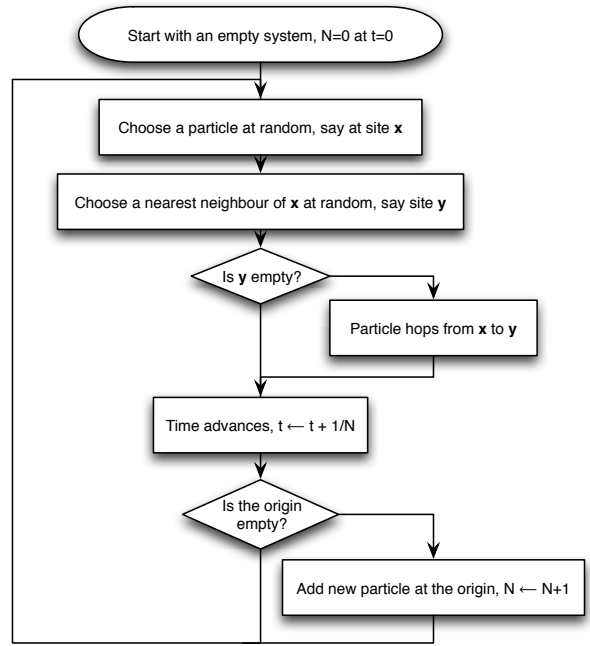}\\
(a)\\
\includegraphics[width=0.5\textwidth]{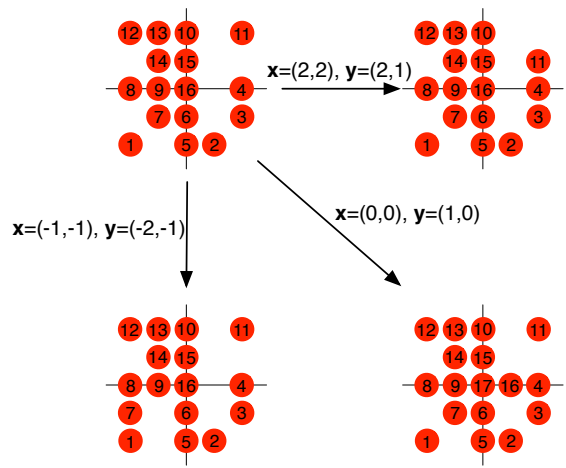}\\
(b)\\
\caption{(a) The rules of the SEP model with the localized source.
(b) Example of state transitions in the model for $d=2$. Top left: a possible configuration of the system after $N=16$
particles have been injected. Particles are numbered 1--16 in the order of arrival; the newest particle
is at the origin. Top right: if particle 11 at ${\bf x}=(2,2)$ and then its nearest neighbour site ${\bf y}=(2,1)$ are randomly chosen, particle 11 hops to $(2,1)$. Bottom left: if particle 7 at ${\bf x}=(-1,-1)$ and then its nearest neighbour site ${\bf y}=(-2,1)$ are randomly chosen, particle 7 hops to $(-2,1)$. If nearest neighbour site
$(0,-1)$ is chosen instead, the particle does not hop because that site is occupied by particle 6.
Bottom right: if particle 16 at the origin is chosen, and then its nearest neighbour site $(1,0)$ is chosen,
the particle hops there, and a new particle 17 is injected at the origin. Not shown: if the wedged particle 9 is chosen,
then regardless of the choice of nearest neighbour, the transition is back into the same configuration. Many other transitions
are also possible. Regardless of the transition chosen, time advances by $\sfrac{1}{16}$.
\label{fig:SSEP-flowchart-and-example-transitions}
}
  \end{center}
\end{figure}

We formalize the system with the following rules.  The system is
initially empty, so that at time $t=0_{+}$ only the origin is
occupied.
\begin{enumerate}
\item One particle, say at site ${\bf x}$, is randomly chosen. 
\item A nearest-neighbour site of ${\bf x}$, say ${\bf y}$, is randomly chosen. 
\item If ${\bf y}$ is empty, the particle hops from ${\bf x}$ to ${\bf y}$; otherwise it remains at ${\bf x}$. 
\item Time advances, $t \leftarrow t+\sfrac{1}{N}$, where $N$ is the current number of particles in the system.
\item If the chosen particle moved from the origin, ${\bf x}={\bf 0}$, a new particle is added at the origin: $N \leftarrow N+1$. 
\item Go back to step 1.
\end{enumerate}
These rules, which also represent the core of our numerical simulations,
are illustrated in Figure~\ref{fig:SSEP-flowchart-and-example-transitions}.

\subsection{Number of particles for the SEP}
We first examine the \emph{number of injected particles}.
We use the notation $N_d\ssep(t)$ for the total number of particles to emphasize the dependence on the spatial dimension $d$. 
The mean $\langle N_d\ssep(t) \rangle$ grows~\cite{Krapivsky2012} as
\begin{equation}
\label{Nav}
 \langle N_d\ssep(t)\rangle \simeq 
\begin{cases}
\sqrt{8t/\pi}                    &d=1\\
\pi\, t/\ln t                       &d=2\\
(2W_d)^{-1}\,t               &d\geq 3
\end{cases}
\end{equation}
Here $W_d$ are the Watson integrals~\cite{Watson1939}
\begin{equation}
\label{Watson_Int}
W_d=\int_0^{2\pi}\ldots\int_0^{2\pi} \frac{1}{Q({\bf q})}\prod_{i=1}^d \frac{dq_i}{2\pi}
\end{equation} 
where ${\bf q} = (q_1,\ldots,q_d)$ and $Q({\bf q}) = \frac{2}{d}\sum_{1\leq i\leq d} (1-\cos q_i)$. 
For the cubic lattice $\mathbb{Z}^3$, the Watson integral  has been expressed~\cite{Glasser1977} via Euler's gamma function
$W_3 = \frac{\sqrt{6}}{64\,\pi^3}\, \Gamma\left(\frac{1}{24}\right)\, \Gamma\left(\frac{5}{24}\right)\, \Gamma\left(\frac{7}{24}\right)\, \Gamma\left(\frac{11}{24}\right) = 0.75819303\ldots$.

Equation \eqref{Nav} shows that the critical
dimension is 2, as for $d>2$ the growth law of $\langle N_d\rangle$ becomes
universal, namely linear in time. In two dimensions, the difference
from the higher-dimensional behaviour is logarithmic, i.e., rather
small. 
We also emphasize that the average total number of injected
particles is lattice-independent when $d\leq 2$. The explanation of
this behaviour is simple: in one and two dimensions, the density varies
on a scale which grows with time, so that the lattice structure is
asymptotically irrelevant. For $d\geq 3$, the results are
lattice-dependent.

\begin{figure}[htbp]
\begin{center}
\includegraphics[width=0.5\textwidth]{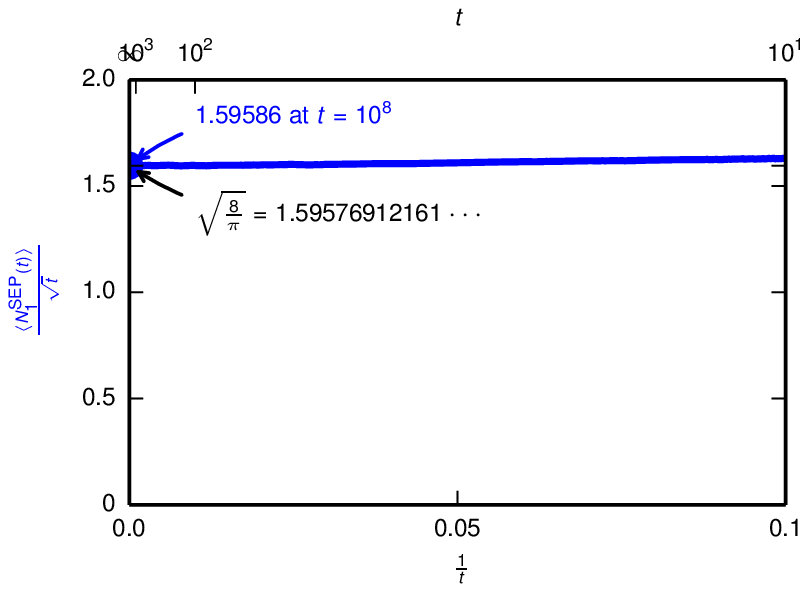}\\
(a)\\
\includegraphics[width=0.5\textwidth]{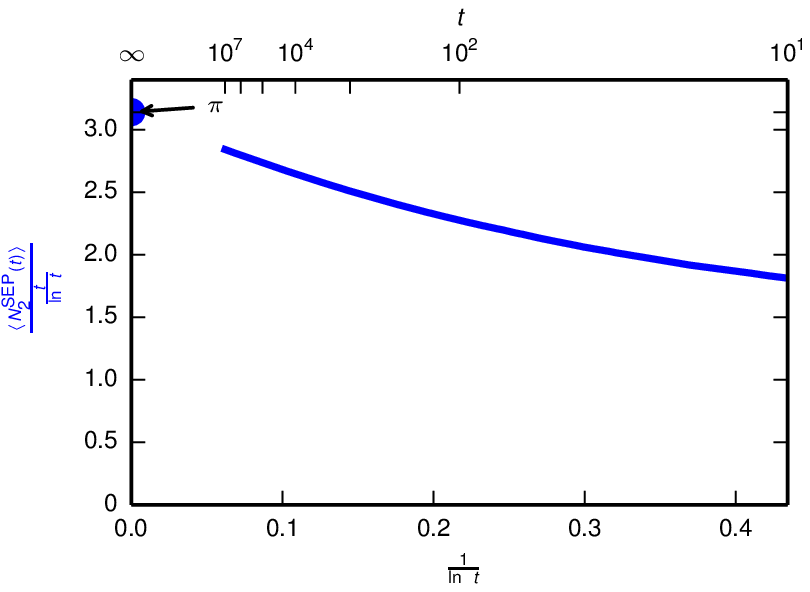}\\
(b)\\
\includegraphics[width=0.5\textwidth]{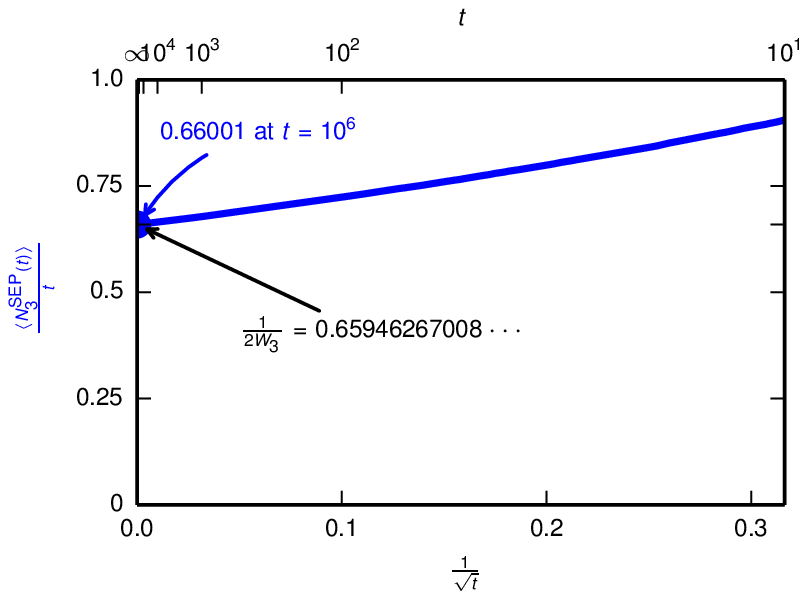}\\
(c)\\
\caption{
Number of injected particles for the SEP: comparison of results from numerical simulations
with theoretical prediction of asymptotic behaviour. 
  (a) One dimension, from 5697 trajectories to $t=10^8$.
  (b) Two dimensions, from 6777 trajectories to $t=10^6$ and 178 to $t=10^7$.
  (c) Three dimensions, from 6594 trajectories to $t=10^5$ and 126 to $t=10^6$.
\label{fig:supplementary-figure-1}
}
\end{center}
\end{figure}

Comparing with simulations,\footnote{Numerical simulations were prototyped in Haskell, and final programs written in C. They were run
on a cluster of workstations with a total of 300 cores over a period of 8 months.}
Figure~\ref{fig:supplementary-figure-1},
in one dimension
there is excellent agreement both for the $\sqrt{t}$ dependence on the time and for the amplitude: 
\begin{equation}
\label{N1:comp}
\frac{\langle N_1\ssep(t)\rangle}{\sqrt{t}}=
\begin{cases}
\sqrt{8/\pi}= 1.595769\ldots  &\text{prediction}\\
\approx 1.59586                     &\text{simulations}
\end{cases}
\end{equation}

\begin{figure}
\begin{center}
\includegraphics[width=0.5\textwidth]{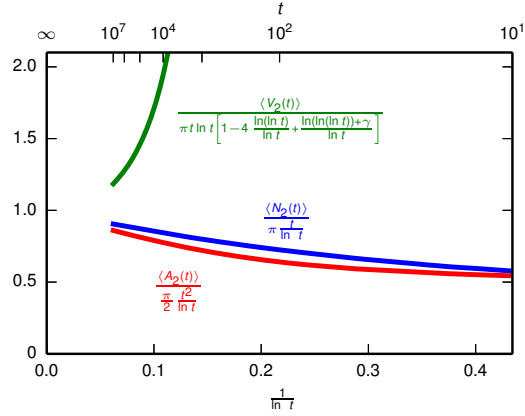}
\caption{Numerical simulations compared with theoretical predictions of asymptotic behaviour for the SEP in two
dimensions, $d=2$. Each curve shows the ratio of the estimate of the mean value of a quantity, from simulations
(6777 trajectories to $t=10^6$ and 178 to $t=10^7$),
and its theoretical prediction, as a function of time. The quantities are,
in blue, the number of particles $N_2(t)$;
in green, the number of unique visited sites $V_2(t)$;
and in red, the activity $A_2(t)$ (i.e., total number of site visits).
Note the scaling of the horizontal axis as $\sfrac{1}{\ln t}$.
As $t \to \infty$ (towards the left edge of the plot), all curves approach 1,
indicating agreement between theoretical predictions and numerical simulations,
even in the difficult-to-analyse case $d=2$. 
Figures~\ref{fig:supplementary-figure-1},~\ref{fig:supplementary-figure-2},and~\ref{fig:supplementary-figure-3}
show details for $d=1,2,3$,
and 
Figures~\ref{fig:supplementary-figure-7}--\ref{fig:supplementary-figure-6}
show analogous results for the model without the exclusion constraint.
\label{fig:SEP-2D-NVA}
}
  \end{center}
\end{figure}

\noindent In two dimensions, a more careful analysis (~\ref{subs}) shows that the convergence to the leading asymptotic given in \eqref{Nav} is very slow:
\begin{equation}
\label{slow}
\langle N_2\rangle \simeq \frac{\pi\, t}{\ln t}\left[1-\frac{\gamma+\ln[\ln(16\sqrt{3}-8)]}{\ln t}+\ldots\right]
\end{equation}
where $\gamma\approx 0.5772$ is Euler's constant. 
The size of our
simulations, or indeed any feasible simulations, is insufficient to extract this behaviour reliably (Figure~\ref{fig:SEP-2D-NVA}).
In three dimensions, from \eqref{Nav},
\begin{equation}
\label{N3:comp}
\frac{\langle N_3\ssep(t)\rangle}{t}=
\begin{cases}
(2W_3)^{-1}= 0.65946\ldots  &\text{prediction}\\
\approx 0.659\ldots                          &\text{simulations}
\end{cases}
\end{equation}
\noindent Thus, in the absence of logarithms simulation results almost perfectly agree 
with theoretical predictions for the asymptotic growth of the number of particles.

\subsection{Number of visited sites for the SEP}
\begin{figure}
\begin{center}
\includegraphics[width=0.5\textwidth]{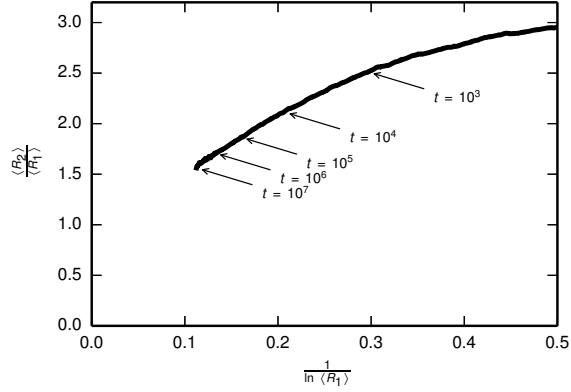}
\caption{The domain of visited sites, $d=2$: Ratio of the smallest circumscribed disc radius to the largest inscribed disc radius, as a function of the latter, from numerical simulations (165 trajectories to $t=10^6$ and 60 to $t=10^7$). Data suggest that $\frac{R_2}{R_1} \sim 1 + \frac{C}{\ln R_1}$, i.e., the relative width of the annulus within which exploration of new territory is occurring decays logarithmically.
\label{fig:annulus-ratio-of-radii}
}
\end{center}
\end{figure}

We now consider the \emph{domain of visited sites}. 
In two dimensions, the visited domains in $\mathbb{Z}^2$, Figure~\ref{fig:domain-example}, 
are fractal-like but remarkably circular. 
Denote by $R_1$, $R_2$ the radius of the largest inscribed and the smallest circumscribed disc, respectively.
Simulations, Figure~\ref{fig:annulus-ratio-of-radii}, suggest that $\frac{R_2}{R_1} \sim 1 + \frac{C}{\ln R_1}$. It would be interesting to explore the shape of the visited domain in detail, e.g., to explain 
this apparent logarithmic decay of the relative width of the annulus, but here we limit ourselves to its volume, the total number of sites each of which has been visited at least once by at least one particle. 
We will show that the average total number of distinct visited sites $ \langle V_d\rangle$ grows as
\begin{equation}
\label{Vav}
 \langle V_d(t)\rangle \propto
\begin{cases}
 t^{d/2} (\ln t)^{d/2}         & d=1, 2, 3\\
 t^2                         & d\geq 4
\end{cases}
\end{equation}
These growth laws indicate that $d^c=4$ plays the role of the upper critical dimension. 

We now provide heuristic arguments in favour of \eqref{Vav}. 
To appreciate the possible time dependence in \eqref{Vav}, 
keep in mind two laws -- the growth law \eqref{Nav} for the average total number of particles, and the well-known growth law~\cite{Weiss1994}
\begin{equation}
\label{distinct}
 \langle \mathcal{V}_d(t)\rangle \sim
\begin{cases}
\sqrt{t}                      &d=1\\
t/\ln t                        &d=2\\
t                               &d>2
\end{cases}
\end{equation}
for the average total number of distinct sited visited by a {\em single} random walker. 

There are two obvious lower bounds, $\langle V_d\rangle > \langle \mathcal{V}_d\rangle$ 
and  $\langle V_d\rangle > \langle N_d\rangle$, which are essentially identical. 
There is also a simple upper bound 
$\langle V_d\rangle < \langle \mathcal{V}_d\rangle \langle N_d\rangle$ 
(because not all particles are introduced at time zero 
and moreover the same site can be visited by different particles). 
Using these bounds and ignoring numerical factors we get 
\begin{equation}
\label{bounds}
\begin{split}
   &\sqrt{t}         <  \langle V_1(t)\rangle < t\\
   &t (\ln t)^{-1}  <  \langle V_2(t)\rangle < t^2 (\ln t)^{-2} \\
   & t                  <  \langle V_d(t)\rangle < t^2, \quad d>2
   \end{split}
\end{equation}

\begin{figure}[htbp]
\begin{center}
\includegraphics[width=0.5\textwidth]{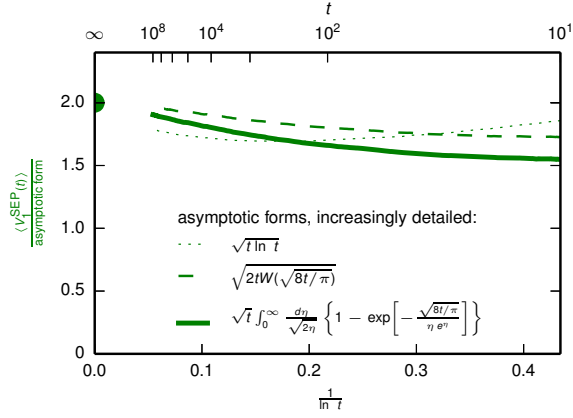}\\
(a)\\
\includegraphics[width=0.5\textwidth]{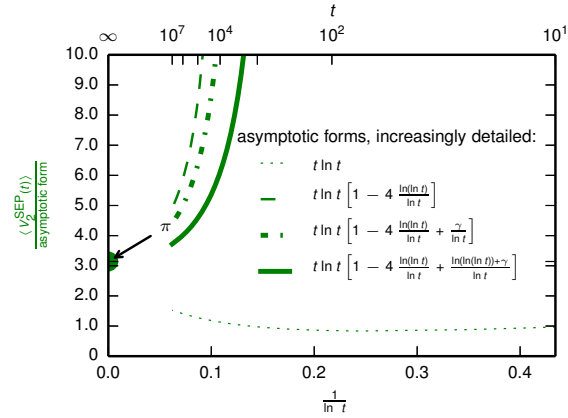}\\
(b)\\
\includegraphics[width=0.5\textwidth]{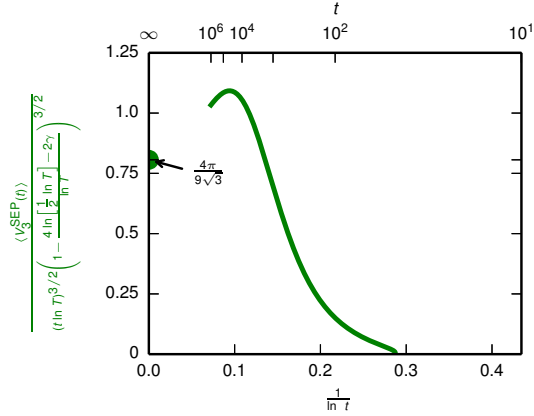}\\
(c)\\
\caption{
Number of visited sites for the SEP: comparison of results from numerical simulations with theoretical prediction of asymptotic behaviour.
 (a) One dimension, from 5697 trajectories to $t=10^8$.
 (b) Two dimensions, from 6777 trajectories to $t=10^6$ and 178 to $t=10^7$.
 (c) Three dimensions, from 6594 trajectories to $t=10^5$ and 126 to $t=10^6$.
\label{fig:supplementary-figure-2}
}
\end{center}
\end{figure}

To obtain stronger heuristic predictions we note that during the time interval $(0, t)$ almost all particles in the system never go more than $\sqrt{t}$ away from the origin. 
This lends support to $\langle V_d(t)\rangle \sim t^{d/2}$. 
This growth law can hold only up to $d=4$, as manifested by the upper bound $\langle V_d(t)\rangle < t^2$. 
These arguments suggest that the algebraic dependence on time is
\begin{equation}
\label{Vav:alg}
 \langle V_d(t)\rangle \propto
\begin{cases}
 t^{d/2}                    &d=1, 2, 3\\
 t^2                         & d\geq 4
\end{cases}
\end{equation}
as in \eqref{Vav}, leaving the possibility of logarithmic corrections. 
Theoretical analyses in dimensions $d=1,2,3$ (Section~\ref{sec:V-derivations}) lead to
the following specific forms (supported by extensive simulations, 
Figure~\ref{fig:supplementary-figure-2}):
\begin{equation}
\label{VavDetailed}
 \langle V_d(t)\rangle \simeq 
\begin{cases}
 2\sqrt{t} \int_0^\infty \frac{d\eta}{\sqrt{2\eta}}
                   \left\{1 - \exp\!\left[-\frac{\sqrt{8t/\pi}}{\eta\,e^\eta}\right]\right\}
                    &d=1\\
\pi t \ln t\left[1-4\,\frac{\ln(\ln t)}{\ln t}+\frac{\gamma}{\ln t}\right]
                    &d=2\\
\frac{4\pi}{9\sqrt{3}}\,(t\ln T)^{3/2}\! \left(1-\frac{4\ln\!\big[\tfrac{1}{2}\ln T\big]-2\gamma}{\ln T}\right)^{3/2},&\\
\textrm{where}\qquad T = t\left[\frac{\ln(2W_3)}{2W_3(2W_3-1)}\right]^2\left(\frac{3}{2\pi}\right)^3
         &d=3\\
\end{cases}
\end{equation}

\subsection{Number of visits for the SEP}
Finally, we consider the
\emph{total number of arrivals} at sites. If the same particle leaves a site and then
returns, a return is counted as a new arrival. Denote by $S_m(t)$ the
total number of sites which have been visited exactly $m$ times during the
time interval $(0,t)$. The zeroth moment of the distribution $S_m(t)$
\begin{equation}
\label{S_zero}
\sum_{m\geq 1} S_m(t) = V(t)
\end{equation}
is merely the total number of visited sites. The first moment of the 
distribution $S_m(t)$ is also interesting: it characterizes the 
integrated ``activity'' of the process, i.e., the total number of arrivals:

\begin{equation}
\label{S_one}
\sum_{m\geq 1} mS_m(t) = A(t)
\end{equation}

\begin{figure}[htbp]
\begin{center}
\includegraphics[width=0.5\textwidth]{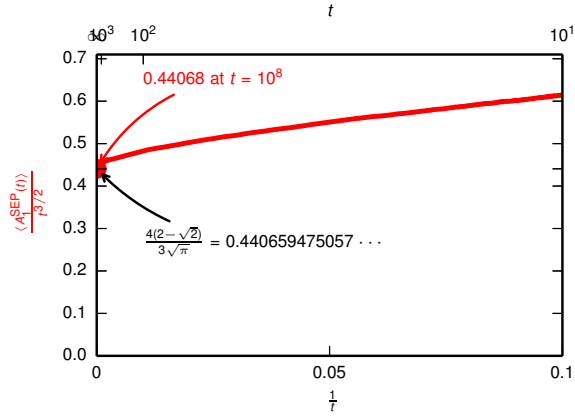}\\
(a)\\
\includegraphics[width=0.5\textwidth]{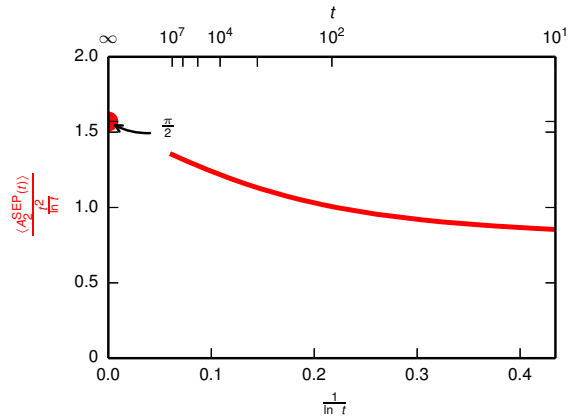}\\
(b)\\
\includegraphics[width=0.5\textwidth]{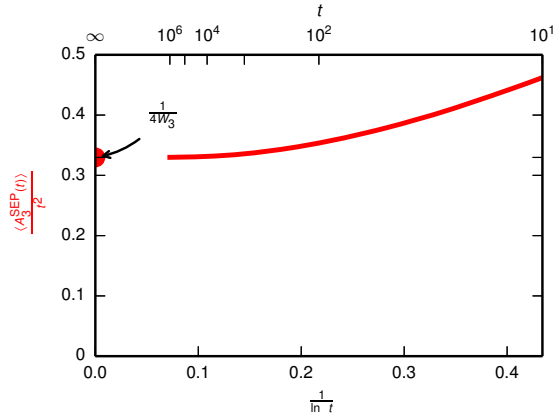}\\
(c)\\
\caption{
Activity (i.e., total number of site visits) for the SEP: comparison of results from numerical simulations with theoretical prediction of asymptotic behaviour.
 (a) One dimension, from 1649 trajectories to $t=10^8$.
 (b) Two dimensions, from 6777 trajectories to $t=10^6$ and 178 to $t=10^7$.
 (c) Three dimensions, from 6594 trajectories to $t=10^5$ and 126 to $t=10^6$.
\label{fig:supplementary-figure-3}
}
\end{center}
\end{figure}

A detailed derivation (Section~\ref{sec:A-derivations}) shows that
\begin{equation}
\label{A:SSEP_num}
 \langle A_d\rangle \simeq
\begin{cases}
 \frac{4(2-\sqrt{2})}{3\sqrt{\pi}}\, t^{3/2}                              & d=1\\
 \tfrac{\pi}{2\ln t}\, t^2                 & d=2 \\
 (4W_d)^{-1}\, t^2                      & d\geq 3
\end{cases}
\end{equation}
as confirmed numerically, Figure~\ref{fig:supplementary-figure-3}.

\section{Random walkers (RW) with a point source}
\label{sec:RW-announce}
In analysing the SEP, we found it convenient
to study in parallel another model, RW,
which is obtained by eliminating the constraint of mutual exclusion
of particles, i.e., a model of quasi-independent random walkers
with a dependence arising only from the localized source.
That model recalls $N$ independent walkers released at once at the origin~\cite{Larralde1992a,Larralde1992b,Sastry1996,Yuste1999,Yuste2000},
but it is simpler, being free of the parameter $N$.
The detailed mathematical development for the asymptotic behaviour of the quantities
$N(t)$, $V(t)$, and $A(t)$ given in the following is in some instances ($N(t)$, $V(t)$) easier
to carry out first for the SEP, in others ($A(t)$) for the RW.
Our mathematical analysis was supported throughout by numerical simulations.

\subsection{The RW model}
The RW model consists of particles undergoing
nearest-neighbour symmetric random walks. Similarly to the case of
the SEP we assume that a new random walker is immediately
deposited at the origin once it becomes empty. Multiple occupancy,
however, is allowed. Despite the lack of direct interaction
between RWs, there is an implicit collective interaction implied by
the input rule: a new RW can be added only when the origin becomes
empty, and this depends on all RWs which were present before the
deposition event. This makes the process involving RWs non-trivial,
and certain features are simpler to compute for the SEP than for the
RWs. This can be appreciated by considering the density at the
origin. In the case of the SEP, $n_{\bf 0}\equiv 1$ since there is
always one particle at the origin. For RWs, the number of particles at
the origin is a random variable, so its full description is provided
by a probability distribution

\begin{equation}
\label{Pk}
P_k(t) :=\text{Prob}[n_{\bf 0}(t)=k]
\end{equation}

For $d=1$ and $d=2$, the probability distribution \eqref{Pk} continues
to evolve and, e.g., the average occupancy of the origin, $\langle
n_{\bf 0}(t)\rangle = \sum_{k\geq 1}kP_k(t)$, grows indefinitely,
albeit anomalously slowly, see Eq.~\eqref{n:origin}. In higher
dimensions, $d\geq 3$, the probability distribution \eqref{Pk} becomes
stationary in the large time limit.

A different version of the model where a large but {\em fixed} number of RWs is
{\em simultaneously} released at a single point has been studied in Refs.~\cite{Larralde1992a,Sastry1996,Yuste1999}. This model has been further analysed and
generalized in subsequent studies, see
e.g., Refs.~\cite{Klafter,Bunde,MT,KMS} and references therein. In our
setting, the number of RWs grows with time, but one can still adopt
the methods of Refs.~\cite{Larralde1992a,Sastry1996,Yuste1999} to investigate, e.g., the
average total number of distinct sites visited by RWs.

The RW model is formalized as follows:
\begin{enumerate}
\item One RW, say at site ${\bf x}$, is randomly chosen, and it hops to a randomly chosen neighbouring site of ${\bf x}$. 
\item Time advances, $t \leftarrow t+\sfrac{1}{N}$, where $N$ is the current number of RWs in the system.
\item If the chosen RW was at the origin and it was the {\em only} particle at the origin,  
a new RW is added at the origin: $N \leftarrow N+1$. 
\item Go back to step 1.
\end{enumerate}

\subsection{Number of particles for the RW model}
Our analytical approach is non-rigorous, but emerging results appear
to be asymptotically exact. We emphasize that, in our set-up, the
addition of the new RW at the origin is ultimately related to the
previous history of the process, so there is an effective
interaction. The analysis is more difficult than in the case of
the SEP where the number of particles at the origin is fixed, $n_{\bf
  0}\equiv 1$.

If the average density $\langle n_{\bf 0}\rangle$ at the origin were known, then
\begin{equation}
\label{nnn}
\langle N_d^\text{RW}\rangle \simeq \langle n_{\bf 0}\rangle \langle N_d\rangle
\end{equation}
with $\langle N_d\rangle$ corresponding to the SEP, where $\langle
n_{\bf 0}\rangle\equiv 1$, and hence given by \eqref{Nav}. Let us
assume that $\langle n_{\bf 0}\rangle$ is a slowly varying function of
time; we will confirm this assumption a posteriori. Thus the
distribution \eqref{Pk} is essentially an equilibrium distribution
with average density $\langle n_{\bf 0}\rangle$. It proves convenient
to consider the case of finite flux, even small flux $F\ll 1$, when
the additions of new particles are rare and the distribution
\eqref{Pk} is the Poisson distribution. In that case we have $P(0,t) =
e^{-\langle n_{\bf 0}\rangle}$, and hence the average total number of
particles increases according to the rate equation
\begin{equation}
\label{nn}
\frac{d}{dt}\langle N_d^\text{RW}\rangle = F e^{-\langle n_{\bf 0}\rangle}
\end{equation}
Using \eqref{nnn} and \eqref{Nav} we get $\frac{d}{dt}\langle
N_1^\text{RW}\rangle \simeq \langle n_{\bf 0}\rangle/\sqrt{\pi t/2}$
in one dimension. Substituting this into \eqref{nn} we obtain $\langle
n_{\bf 0}\rangle\simeq \frac{1}{2}\ln t$. The leading asymptotic does
{\em not} depend on the flux, and hence we anticipate that the
prediction for $\langle n_{\bf 0}\rangle$ remains correct for large
flux, and in the most interesting case of the infinite flux (into the
empty origin). A more accurate derivation for the case of the infinite
flux 
(Section~\ref{nu_d:RW})
confirms the $\langle n_{\bf
  0}\rangle\simeq \frac{1}{2}\ln t$ asymptotic.

In two dimensions we similarly find
\begin{equation*}
\frac{d}{dt}\langle N_2^\text{RW}\rangle\simeq \langle n_{\bf 0}\rangle\,\frac{\pi}{\ln t}\sim e^{-\langle n_{\bf 0}\rangle}
\end{equation*}
from which $\langle n_{\bf 0}\rangle\simeq \ln(\ln t)$.

\begin{figure}[htbp]
\begin{center}
\includegraphics[width=0.5\textwidth]{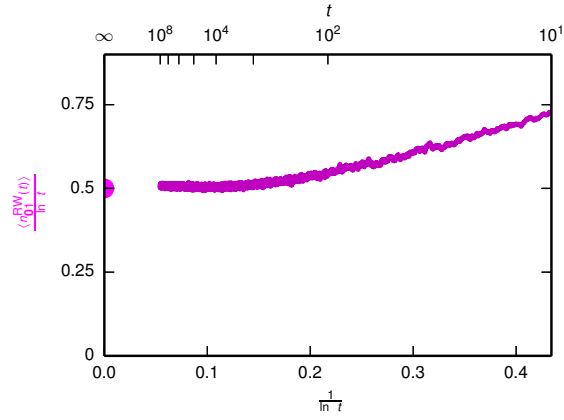}\\
(a)\\
\includegraphics[width=0.5\textwidth]{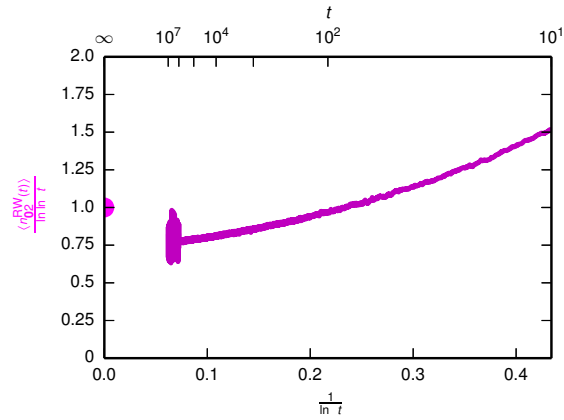}\\
(b)\\
\includegraphics[width=0.5\textwidth]{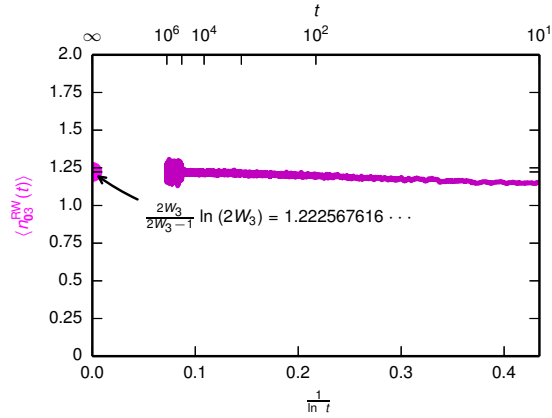}\\
(c)\\
\caption{
Origin occupancy for RWs: comparison of results from numerical simulations with theoretical prediction of asymptotic behaviour.
 (a) One dimension, from 3315 trajectories to $t=10^8$.
 (b) Two dimensions, from 9008 trajectories to $t=10^6$ and 74 to $t=10^7$.
 (c) Three dimensions, from 6689 trajectories to $t=10^5$ and 261 to $t=10^6$.
\label{fig:supplementary-figure-7}
}
\end{center}
\end{figure}

When $d\geq 3$, the average density at the origin approaches $2W_d [2W_d-1]^{-1} \ln(2W_d)$, as we show 
in Section~\ref{nu_d:RW}.
Thus the average density at the origin reads
\begin{equation}
\label{n:origin}
\langle n_{{\bf 0}d}\rangle \simeq
\begin{cases}
\frac{1}{2} \ln t                                  &d=1\\
\ln(\ln t)                                             &d=2\\
2W_d [2W_d-1]^{-1} \ln(2W_d)        &d\geq 3
\end{cases}
\end{equation}
For the cubic lattice, $\langle n_{{\bf 0}3}\rangle=1.222567616\ldots$; see 
Figure~\ref{fig:supplementary-figure-7}.
The average total number of particles grows as
\begin{equation}
\label{Nav:RW}
 \langle N_d\rangle \simeq
\begin{cases}
\sqrt{2t/\pi}\, \ln t                             &d=1\\
\pi t\,\frac{\ln\ln t}{\ln t}                    &d=2\\
[2W_d-1]^{-1} \ln(2W_d)\,\, t          &d\geq 3
\end{cases}
\end{equation}
Hereinafter we write $\langle N_d\rangle$ instead of $\langle
N_d^\text{RW}\rangle$ when there is no danger of misinterpretation.

In one and two dimensions, the sub-leading terms are only formally
negligible; in practice they are almost as large as the leading
terms. Indeed, repeating the above analysis and trying to keep sub-leading
terms, one gets
\begin{equation}
\label{n1:origin}
\langle n_{\bf 0}\rangle = \tfrac{1}{2} \ln t + C_2\ln\ln t + \ldots
\end{equation}
in one dimension. The amplitude $C_2$ is very difficult to compute,
and even if we were able to compute it, the following sub-sub-leading
term which is not displayed in \eqref{n1:origin} will involve an even
nastier repeated logarithm: $\ln \ln\ln t$. Ignoring these
difficult-to-compute corrections, we get surprisingly good
agreement:
\begin{equation}
\label{N1:comp_RW}
\frac{\langle N_1\rangle}{\sqrt{t}\,\ln t}=
\begin{cases}
\sqrt{2/\pi}=0.79788\ldots      &\text{prediction}\\
\approx 0.77                          &\text{simulations}
\end{cases}
\end{equation}

Similarly in two dimensions
\begin{equation}
\label{n2:origin}
\langle n_{\bf 0}\rangle = \ln\ln t + C_3 \ln\ln\ln t + \ldots
\end{equation}
with unknown amplitude $C_3$. The appearance of repeated logarithms
makes it doubtful that one can confirm heuristic predictions for the
leading terms.

\begin{figure}[htbp]
\begin{center}
\includegraphics[width=0.5\textwidth]{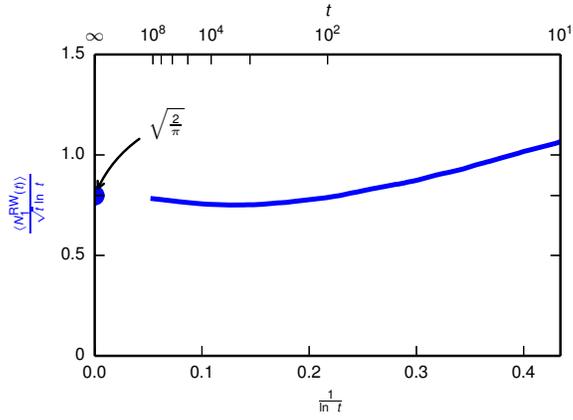}\\
(a)\\
\includegraphics[width=0.5\textwidth]{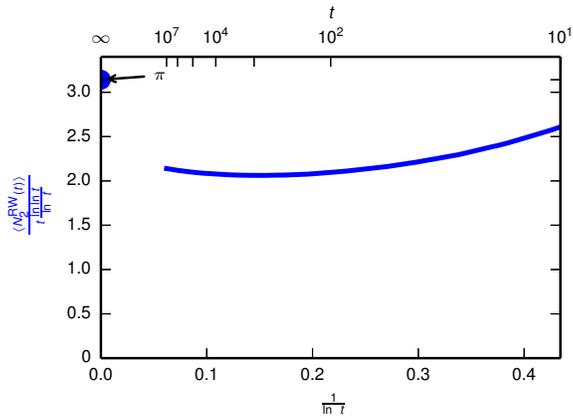}\\
(b)\\
\includegraphics[width=0.5\textwidth]{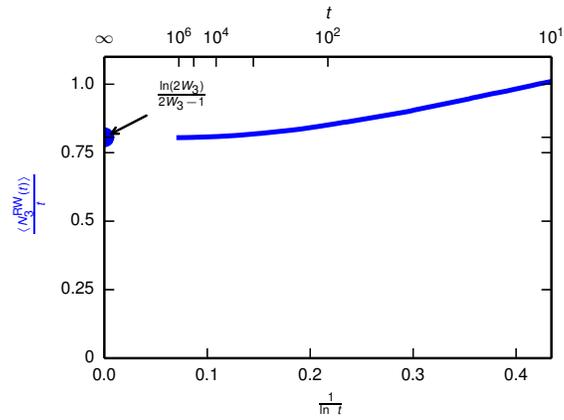}\\
(c)\\
\caption{
Number of injected particles for RWs: comparison of results from numerical simulations
with theoretical prediction of asymptotic behaviour. 
  (a) One dimension, from 3315 trajectories to $t=10^8$.
  (b) Two dimensions, from 9008 trajectories to $t=10^6$ and 74 to $t=10^7$.
  (c) Three dimensions, from 6689 trajectories to $t=10^5$ and 261 to $t=10^6$.
\label{fig:supplementary-figure-4}
}
\end{center}
\end{figure}

In three dimensions, the linear growth with time is in excellent
agreement with simulation results. As for the amplitude, the
theoretical prediction is
\begin{equation}
\label{N3:comp_RW}
\frac{\langle N_3\rangle}{t} = \frac{\ln(2W_3)}{2W_3-1}=0.806237705\ldots
\end{equation}
Numerically the amplitude is approximately $0.81$;
see 
Figure~\ref{fig:supplementary-figure-4}.

\subsection{Density distribution at the origin}
\label{nu_d:RW}

The probability distribution \eqref{Pk} describing RWs at the origin satisfies 
\begin{equation}
\label{Pk2}
\frac{dP_k}{dt}=(k+1)P_{k+1}-(k+\mu)P_k+\mu P_{k-1}
\end{equation}
for $ k\geq 2$ and
\begin{equation}
\label{Pk1}
\frac{dP_1}{dt} = 2P_2-\mu P_1
\end{equation}
Here $\mu$ is the average density on the sites neighbouring the
origin. In the long time limit we can neglect the terms on the
left-hand side of Eqs.~\eqref{Pk2}--\eqref{Pk1}. This is obvious when
$d\geq 3$, since in this case the probability distribution reaches a
stationary state. When $d=1$ or $d=2$, the left-hand sides can be
ignored in the realm of a quasi-stationary approximation; one can make
such an assumption, find a solution, and justify the quasi-stationary
approximation a posteriori.

Solving the stationary version of \eqref{Pk1} and then the following
equations \eqref{Pk2} we find $P_k=P_1\,\mu^{k-1}/k!$ and fix $P_1$
through the normalization $\sum_{k\geq 1}P_k=1$ to yield
\begin{equation}
\label{Pk_mu}
P_k=\frac{\mu^{k}}{k!}\,\frac{1}{e^\mu -1}
\end{equation}

For $d\geq 3$, we have
\begin{equation}
\label{N_RW:eq}
\frac{d}{dt}\langle\, N_d\rangle = P_1 = \frac{\mu}{e^\mu -1}
\end{equation}
Using \eqref{Pk_mu} we find that the average density at the origin 
\begin{equation}
\label{n_mu}
\langle n_{\bf 0}\rangle = \sum_{k\geq 1}kP_k = \mu\,\frac{e^\mu}{e^\mu-1}
\end{equation}

Recall that $n_{\bf 0}/\mu = 2W_d/(2W_d-1)$ in the case of the SEP,
see Ref.~\cite{Krapivsky1992}. The ratio is the same in the case of RWs (as the
governing equations for the average densities are
identical). Therefore \eqref{n_mu} gives
\begin{equation}
\label{Wd_mu}
\frac{2W_d}{2W_d-1} = \frac{e^\mu}{e^\mu-1}
\end{equation}
from which $\mu=\ln(2W_d)$. Substituting this into \eqref{n_mu} we
establish the announced result \eqref{n:origin}.

For $d=1$ and $d=2$, the average density at the origin $\langle n_{\bf
  0}\rangle$ and the average density $\mu$ on the sites neighbouring
the origin both diverge in the long time limit, and asymptotically
$\langle n_{\bf 0}\rangle\simeq \mu$ due to \eqref{n_mu}. Hence
\eqref{N_RW:eq} becomes
\begin{equation}
\label{N_RW:12}
\frac{d}{dt}\langle\, N_d\rangle = \langle n_{\bf 0}\rangle\,
e^{-\langle n_{\bf 0}\rangle}
\end{equation}
in the leading order. This is a more precise equation than \eqref{nn},
yet it yields the same leading asymptotic.

\section{The volume of the domain of visited sites}
\label{sec:V-derivations}
For RWs, we shall present theoretical evidence in favour of the growth
laws \eqref{Vav} and show that in the physically relevant
dimensions the amplitudes read
\begin{equation}
\label{C123}
C_1 = 2, \quad C_2 = \pi, \quad C_3 = \frac{4\pi}{9\sqrt{3}}
\end{equation}

\begin{figure}[htbp]
\begin{center}
\includegraphics[width=0.5\textwidth]{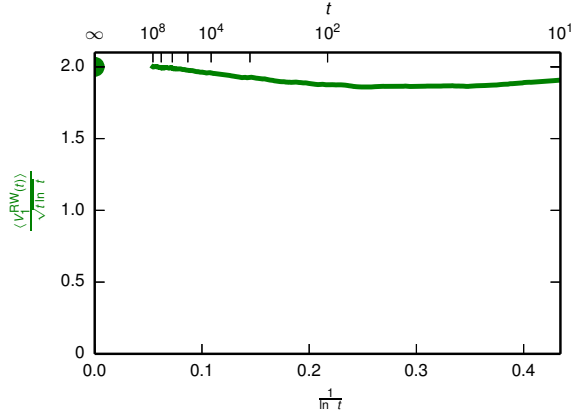}\\
(a)\\
\includegraphics[width=0.5\textwidth]{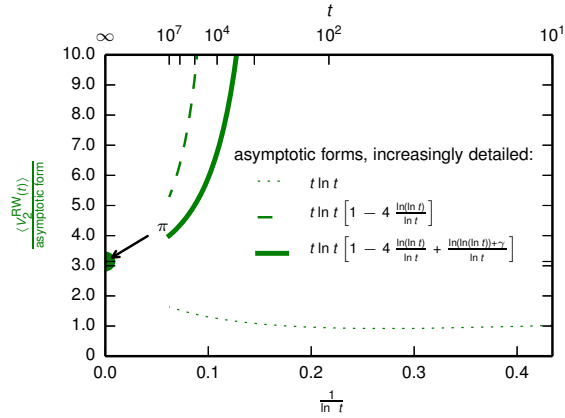}\\
(b)\\
\includegraphics[width=0.5\textwidth]{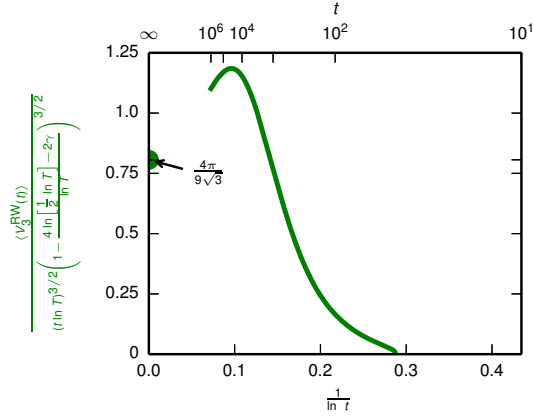}\\
(c)\\
\caption{
Number of visited sites for RWs: comparison of results from numerical simulations 
with theoretical prediction of asymptotic behaviour.
 (a) One dimension, from 3315 trajectories to $t=10^8$.
 (b) Two dimensions, from 9008 trajectories to $t=10^6$ and 74 to $t=10^7$.
 (c) Three dimensions, from 6689 trajectories to $t=10^5$ and 261 to $t=10^6$.
\label{fig:supplementary-figure-5}
}
\end{center}
\end{figure}

Our simulations, Figures~\ref{fig:supplementary-figure-2}~and~\ref{fig:supplementary-figure-5},
indicate that the growth laws \eqref{Vav} apply both
to the SEP and RWs. The amplitudes $C_d$ observed in simulations are
larger for RWs than for the SEP. We believe that the amplitudes
are the same in the physically relevant dimensions and the discrepancy
is caused by large sub-leading corrections. 
In Section~\ref{inf:1d}
we show that $C_1 = 2$ for both the SEP and RWs; we also estimate the
ratio of effective amplitudes and show that it exhibits an anomalously
slow convergence to unity:
\begin{equation}
\label{A1_ratio}
\frac{C_1^\text{RW}}{C_1^\text{SEP}} -1 
\simeq  \frac{\ln\!\big(\tfrac{1}{2}\ln t\big)}{\ln t}
\end{equation}
In two dimensions, we shall argue [cf. \eqref{V2} and \eqref{V2:SSEP}] that
\begin{equation*}
\frac{C_2^\text{RW}}{C_2^\text{SEP}} -1 
\simeq  \frac{\ln(\ln(\ln t))}{\ln t}
\end{equation*}
We have not computed the amplitudes in higher than three
dimensions. Intuitively, one anticipates that for $d\geq 4$ the
amplitudes $C_d$ remain larger for RWs than for the SEP even in
the $t\to\infty$ limit.

\subsection{The volume of the domain of visited sites: one dimension}
\label{inf:1d}

In one dimension, it suffices to study the one-sided problem. We wish to
compute the probability $\Pi(L,t)$ that the particles never went beyond
distance $L$ during the time interval $(0,t)$. To this end we put an
artificial boundary at $x=L$. Mathematically, we must solve the
diffusion equation
\begin{equation}
\label{diff}
\frac{\partial \rho}{\partial t} = \frac{1}{2}\,\frac{\partial^2 \rho}{\partial x^2}
\end{equation}
on the interval $0\leq x\leq L$. The initial condition is
\begin{equation}
\label{IC_true}
\rho(x, t=0) = 0
\end{equation}
and the absorbing boundary condition
\begin{equation}
\label{BC_L}
\rho(x=L, t) = 0
\end{equation}
For the SEP, there is another boundary condition
\begin{equation}
\label{BC_0}
\rho(x=0, t) = 1
\end{equation}
modelling the source. The probability $\Pi(L,t)$ is then 
\begin{equation}
\Pi(L,t) = \exp\!\left[\int_0^t dT\,\frac{\partial \rho(L, T)}{\partial x}\right]
\end{equation}
The probability that the distance is exactly $L$ is given by $\frac{\partial \Pi}{\partial L}$, and the average distance
\begin{equation}
\label{Lav}
\langle L\rangle = \int_0^\infty dL\,L\,\frac{\partial \Pi}{\partial L} =  \int_0^\infty dL\,[1-\Pi(L,t)]
\end{equation}

The main contribution is gathered in the region 
\begin{equation}
\label{L_large}
L\gg \sqrt{t}
\end{equation}
We assume this to hold and justify a posteriori. Using \eqref{L_large}
we can simplify the problem, namely we can consider \eqref{diff} on
the real line $-\infty<x<\infty$ subject to the initial condition
\begin{equation}
\label{IC}
\rho(x, t=0) = 
\begin{cases}
2  & x<0 \\
0  & 0<x<2L \\
-2 & x>2L
\end{cases}
\end{equation}
Then the absorbing boundary condition \eqref{BC_L} is obeyed due to
symmetry, while the boundary condition \eqref{BC_0} is valid thanks to
\eqref{L_large}. Solving \eqref{diff} subject to \eqref{IC} yields
\begin{equation*}
\rho(x, T) = \sqrt{\frac{2}{\pi T}}\left[\int_0^\infty dy\,
e^{-\frac{(x+y)^2}{2T}}-\int_{2L}^\infty dy\,e^{-\frac{(y-x)^2}{2T}}\right]
\end{equation*}
{}from which
\begin{equation}
\frac{\partial \rho(x=L, T)}{\partial x} = -\sqrt{\frac{8}{\pi T}}\, \exp\!\left(-\frac{L^2}{2T}\right)
\end{equation}
and therefore
\begin{equation}
\label{PLt}
\Pi(L,t) = \exp\!\left[-\int_0^t dT\,\sqrt{\frac{8}{\pi T}}\, \exp\!\left(-\frac{L^2}{2T}\right)\right]
\end{equation}
Writing 
\begin{equation}
\label{Lt}
T = t\tau, \quad L = \sqrt{2t\eta}
\end{equation}
we recast \eqref{PLt} into
\begin{eqnarray}
\label{PL}
\Pi(L,t) &=& \exp\!\left[-\sqrt{\frac{8t}{\pi}}\int_0^1 \frac{d\tau}{\sqrt{\tau}}\,e^{-\eta/\tau}\right]\nonumber\\
           &=& \exp\!\left[-\sqrt{\frac{8t}{\pi}}\,\frac{e^{-\eta}}{\eta}\right]
\end{eqnarray}
where the second line is the leading asymptotic which applies when
$\eta\gg 1$. Substituting \eqref{Lt}--\eqref{PL} into \eqref{Lav} we
compute $\langle L\rangle$ and determine $\langle V_1\rangle =
2\langle L\rangle + 1$
\begin{equation}
\label{V1_int}
\langle V_1\rangle  =  2\sqrt{t} \int_0^\infty \frac{d\eta}{\sqrt{2\eta}}
                   \left\{1 - \exp\!\left[-\frac{\sqrt{8t/\pi}}{\eta\,e^\eta}\right]\right\}
\end{equation}
In the large time limit, the integral in \eqref{V1_int} approaches $\sqrt{2\zeta}$, where $\zeta$ is a root of 
\begin{equation}
\label{zeta}
\zeta\,e^\zeta=\sqrt{8t/\pi}
\end{equation}
Thus
\begin{equation}
\label{V1_simple}                   
\langle V_1\rangle  \simeq  2\sqrt{2\zeta t}
\end{equation}
A root of Eq.~\eqref{zeta} is known as the Lambert $W$ function:
$\zeta=W(\sqrt{8t/\pi})$. Asymptotically $\zeta\simeq \tfrac{1}{2}\ln
t$, and therefore the average total number of visited sites grows as
\begin{equation}
\label{V1}
\langle V_1\rangle \simeq 2\sqrt{t\,\ln t}
\end{equation}
in agreement with \eqref{Vav}. [Equation \eqref{V1} also gives the
  announced result \eqref{C123} for the amplitude in one dimension.]
The asymptotic growth \eqref{V1} is faster than $\sqrt{t}$. This
justifies the assumption \eqref{L_large} which was used in the above
analysis.

It is difficult to confirm the amplitude $C_1=2$ numerically. The
solution to Eq.~\eqref{zeta} actually reads $\zeta\simeq
\tfrac{1}{2}\ln t - \ln(\ln t)+\ldots$, so the approach to the leading
asymptotic growth is very slow. One can take \eqref{V1_int} as the
theoretical prediction, compute the integral numerically as a function of
time, and compare the outcome with simulations, 
Figure~\ref{fig:supplementary-figure-1}a.

For RWs, the density is obtained by multiplying the density $\rho(x,
T)$ corresponding to the SEP problem by the average density
$\tfrac{1}{2}\ln T$ of RWs at the origin. Thus instead of \eqref{PLt}
we obtain
\begin{equation*}
\Pi(L,t) = \exp\!\left[-\int_0^t dT\,\sqrt{\frac{2}{\pi T}}\,\ln T\, \exp\!\left(-\frac{L^2}{2T}\right)\right]
\end{equation*}
and instead of \eqref{PL} we get
\begin{equation*}
\Pi(L,t) = \exp\!\left[-\ln t\,\sqrt{\frac{2t}{\pi}}\,\frac{e^{-\eta}}{\eta}\right]
\end{equation*}
from which
\begin{equation}
\label{V1_int:RW}
\langle V_1\rangle =  2\sqrt{t} \int_0^\infty \frac{d\eta}{\sqrt{2\eta}}
                   \left\{1 - \exp\!\left[-\frac{\sqrt{2t/\pi}\,\ln t}{\eta\,e^\eta}\right]\right\}
\end{equation}

The average number of visited sites is given by the same formula \eqref{V1_simple} as before, where $\zeta$ is now a root of 
\begin{equation}
\label{zeta:RW}
\zeta\,e^\zeta=\sqrt{\frac{2t}{\pi}}\, \ln t
\end{equation}
which can be written through the Lambert $W$ function:
$\zeta=W(\sqrt{2t/\pi}\, \ln t)$.  The leading asymptotic is the same
as in the case of the SEP, $\zeta\simeq \tfrac{1}{2}\ln t$, and
therefore the asymptotic growth is given by the same formula
\eqref{V1}. A better approximation is probably provided by
\eqref{V1_int:RW}; see 
Figure~\ref{fig:supplementary-figure-4}a.

Let us try to estimate the discrepancy between the growth of the
average total number of visited sites in the case of the SEP and
RWs. We have
\begin{equation*}
\frac{\langle V_1^\text{RW}\rangle}{\langle V_1\rangle}\simeq \sqrt{\frac{\zeta^\text{RW}}{\zeta}}
\simeq 1+ \frac{\zeta^\text{RW}-\zeta}{2\zeta}
\end{equation*}
where $\zeta^\text{RW}$ is the solution of \eqref{zeta:RW}. Dividing
\eqref{zeta:RW} by \eqref{zeta} we obtain $\zeta^\text{RW}-\zeta\simeq
\ln\!\big(\tfrac{1}{2}\ln t\big)$, and therefore
\begin{equation}
\frac{\langle V_1^\text{RW}\rangle}{\langle V_1\rangle} - 1 
\simeq \frac{\ln\!\big(\tfrac{1}{2}\ln t\big)}{\ln t}
\end{equation}
leading to the announced result \eqref{A1_ratio} for the ratio of effective amplitudes.

\subsection{The volume of the domain of visited sites: higher dimensions}

Consider first a single RW released at the origin at time $t=0$. 
The probability that it will visit site ${\bf x}$ during the time interval $(0,t)$ is 
\begin{equation}
\label{Pxt}
P({\bf x}, t) = \frac{1}{2W_d}\int_0^t \frac{d\tau}{(4\pi D\tau)^{d/2}}\,\exp\!\left[-\frac{x^2}{4D\tau}\right]
\end{equation}
where $x^2=|{\bf x}|^2$ and $D=(2d)^{-1}$ with our choice of the
hopping rates. Equation \eqref{Pxt} is (asymptotically) exact in the
long-time limit, $t\gg 1$, when we can ignore the lattice structure
and use the prediction $(4\pi D\tau)^{-d/2} e^{-x^2/4D\tau}$ from the
continuous approach for the probability to visit site ${\bf x}$ the
last time at time $\tau$ before $t$. We then multiply this probability
by the persistence probability that the RW does not return to
${\bf x}$ during the time interval $(\tau, t)$ and take into account
that this persistence probability saturates at $(2W_d)^{-1}$ when
$d>2$.

When $d>2$, the renormalized flux is finite, so for computing the
average number of visits it suffices to assume that RWs are added at a
constant rate $\Phi_d$; according to \eqref{Nav:RW} the flux is
$\Phi_d=[2W_d-1]^{-1} \ln(2W_d)$, although the actual value will play
a minor role. Thus we release particles at times
$0=t_1<t_2<\cdots<t_N=t$ with $N=\Phi_d t$. The probability that none
of them will visit site ${\bf x}$ is given by $\prod_{1\leq j\leq N}
(1-P_j)$, so the probability that at least one RW will visit site
${\bf x}$ is $1-\prod_{1\leq j\leq N} (1-P_j)$, and the average number
of visited sites is
\begin{equation}
\label{V_integral}
\langle V_d\rangle = \int d{\bf x} \left[1-\prod_{j=1}^N (1-P_j)\right]
\end{equation}
Using \eqref{Pxt} we get 
\begin{equation}
\label{PNj}
P_{N-j} =  \frac{1}{2W_d}\int_0^{tj/N} 
\frac{d\tau}{(4\pi D\tau)^{d/2}}\,\exp\!\left[-\frac{x^2}{4D\tau}\right]
\end{equation}
We will see that the dominant contribution to the integral in
\eqref{V_integral} is gathered in the region where
$\xi=\frac{x^2}{4Dt}\gg 1$. In this region \eqref{PNj} simplifies to
\begin{equation}
\label{PNj:simple}
P_{N-j} =  
\frac{(tj/N)^{2-d/2}}{2W_d(4\pi D)^{d/2}}\,  t^{-1}\xi^{-1}\exp\!\left[-\xi\,\frac{N}{j}\right]
\end{equation}
We now write
\begin{equation*}
\ln \prod_{j=1}^N (1-P_j) 
=\sum_{j=1}^N \ln(1-P_j) \simeq -\int_1^N dj\,P_j 
\end{equation*}
Combining this with \eqref{PNj:simple} and computing the asymptotic behaviour of the integral we arrive at
\begin{equation}
\label{V_int}
\langle V_d\rangle = \frac{\Omega_d (4Dt)^\delta}{2}\int_0^\infty \frac{d\xi}{\xi^{1-\delta}}\!
\left(1-\exp\!\left[-\frac{\Psi_d t^{2-\delta}}{\xi^2 e^\xi}\right]\right)
\end{equation}
where $\Omega_d=2\pi^\delta/\Gamma(\delta)$ is the volume of the
$(d-1)$-dimensional unit sphere,
$\Psi_d=(2W_d)^{-1}\Phi_d(\delta/\pi)^\delta$, and we use the
shorthand notation $\delta=d/2$.  The expression in the brackets in
the integral in \eqref{V_int} is very close to 1 when $\xi<\xi_*$ and
it quickly vanishes when $\xi>\xi_*$, where $\xi_*$ is
determined from
\begin{equation}
\label{xi_*}
\xi_*^2\,e^{\xi_*} = \Psi_d\, t^{2-\delta}
\end{equation}
The integral is $\int_0^{\xi_*} d\xi\,\xi^{\delta-1} =
\delta^{-1}\xi_*^\delta$ in the leading order. {}From \eqref{xi_*} we
see that $\xi_*\simeq (2-\delta)\ln t$, so it diverges when
$\delta<2$, i.e. $d<4$. The above analysis assumes that the major
contribution to the integrals is gathered when $\xi$ is large, and
hence it is justified when $d<4$. Since we also assumed that $d>2$,
the results essentially apply only to $d=3$. Specializing to $d=3$ we
arrive at
\begin{equation}
\label{V3}
\langle V_3\rangle = \frac{4\pi}{9\sqrt{3}}\,(t\ln T)^{d/2}\! \left(1-\frac{4\ln\!\big[\tfrac{1}{2}\ln T\big]-2\gamma}{\ln T}\right)^{3/2}
\end{equation}
Here instead of $\xi_*\simeq \tfrac{1}{2}\ln t$ we used a more precise solution of \eqref{xi_*} in three dimensions:
\begin{equation*}
\xi_*\simeq \tfrac{1}{2}\ln T - 2\ln\!\big[\tfrac{1}{2}\ln T\big]
\end{equation*}
and a more precise expression,
viz. $\tfrac{2}{3}\xi_*^{3/2}+\gamma\sqrt{\xi_*}$ where $\gamma
\approx 0.5772$ is the Euler constant, of the integral in
\eqref{V_int}. We also inserted the constant $\Psi_3^2$ into the time
variable, namely inside the logarithms the time variable is
\begin{equation*}
T = t\left[\frac{\ln(2W_3)}{2W_3(2W_3-1)}\right]^2\left(\frac{3}{2\pi}\right)^3
\end{equation*}

When $d=4$, there is no sharp crossover, and $\langle V_4\rangle \sim
t^2$ without logarithmic corrections. When $d>4$, we have $\langle
V_d\rangle \sim t^2$ as we already explained above, namely because
both the lower and upper bounds scale as $t^2$.

The two-dimensional case is more subtle. We can still use \eqref{Pxt},
but the persistence probability now vanishes: $P_2\simeq 1/\ln t$. It
vanishes so slowly that we can still use the same quantity $1/\ln t$
independently of the time when the RW was released. We can also ignore
the fact that the flux slowly varies with time and merely use $N=\pi
t\,\frac{\ln\ln t}{\ln t}$, see \eqref{Nav:RW}. Repeating the above
analysis we get
\begin{equation}
\label{V_2}
\langle V_2\rangle = \pi t\int_0^\infty d\xi
\left(1-\exp\!\left[-\frac{\ln\ln t}{(\ln t)^2}\,\frac{t}{\xi^2\,e^\xi}\right]\right)
\end{equation}
The crossover occurs around $\xi_*$ which is implicitly determined by 
\begin{equation*}
\xi_*^2\,e^{\xi_*} = t\,\frac{\ln\ln t}{(\ln t)^2}
\end{equation*}
{}from which $\xi_*\simeq \ln t - 4\ln(\ln t) + \ln(\ln(\ln t))$. The 
integral in \eqref{V_2} is $\xi_*$ in the leading order. A more precise 
estimate of the integral is given by
\begin{equation*}
\int_0^\infty d\xi
\left(1-\exp\!\left[-\frac{\xi_*^2\,e^{\xi_*}}{\xi^2\,e^\xi}\right]\right) = \xi_*+ \gamma
\end{equation*}
The constant is Euler's constant due to the identity
\begin{equation*}
\int_0^\infty dy\,
\big(1-e^{-e^{-y}}-e^{-e^{y}}\big) = \gamma
\end{equation*}
Therefore 
\begin{equation}
\label{V2}
\langle V_2\rangle \simeq \pi t \ln t\left[1-4\,\frac{\ln(\ln t)}{\ln t}+\frac{\ln(\ln(\ln t))+\gamma}{\ln t}\right]
\end{equation}

The growth laws \eqref{V3} and \eqref{V2} hint why it is so 
difficult to extract the true asymptotic behaviours from numerics;
see 
Figure~\ref{fig:supplementary-figure-2}.

In physically relevant dimensions, the main contribution to the total
number of distinct visited sites is gathered in the region
$\xi=\frac{x^2}{4Dt}\gg 1$ where the density is small and the
difference between RWs and the SEP is negligible. Therefore the
leading asymptotic behaviour of $\langle V_d\rangle$ is expected to be
the same when $d=1,2,3$. 
Above, this was analysed in more detail for the $d=1$ case.
In two dimensions, one anticipates that
\begin{equation*}
\langle V_2^\text{SEP}\rangle = \pi t\int_0^\infty d\xi
\left(1-\exp\!\left[-\frac{1}{(\ln t)^2}\,\frac{t}{\xi^2\,e^\xi}\right]\right)
\end{equation*}
leading to
\begin{equation}
\label{V2:SSEP}
\langle V_2^\text{SEP}\rangle \simeq \pi t \ln t\left[1-4\,\frac{\ln(\ln t)}{\ln t}+\frac{\gamma}{\ln t}\right]
\end{equation}

In addition to the average number of visited sites, one would like to
compute the full probability distribution. This is a much more
challenging problem, especially when $d>1$. Even in one dimension when
the visited domain is an interval, the problem is rather challenging
and it has been studied only in the situation \cite{Larralde1992a,KMS} when the
total number of random walks is fixed and they were all simultaneously
released. Even in this situation, conditioning the trajectories of the
RWs to a given number of visited sites introduces effective
correlations between the walkers~\cite{KMS}. In our setting, when the
RWs are released throughout the evolution, there are additional
correlations which are built to allow injection events, there are
fluctuations in the total number of injected RWs, etc. Let us make a
bold assumption that these additional effects do not change the
behaviour in the leading order.  Specializing the results of
Ref.~\cite{KMS} to our setting yields the probability distribution
\begin{equation}
P(V_1,t) = \sqrt{\tfrac{\ln t}{t}}\,\mathcal{D}(s), \quad s = \sqrt{\tfrac{\ln t}{t}}\,[V_1-\langle V_1\rangle]
\end{equation}
with the scaled probability distribution given by
\begin{equation*}
\mathcal{D}(s) = 2e^{-s}K_0(2e^{-s/2})
\end{equation*}
where $K_0$ is the modified Bessel function.

\section{Statistics of visits to sites}
\label{sec:A-derivations}

\begin{figure}[htbp]
\begin{center}
\includegraphics[width=0.5\textwidth]{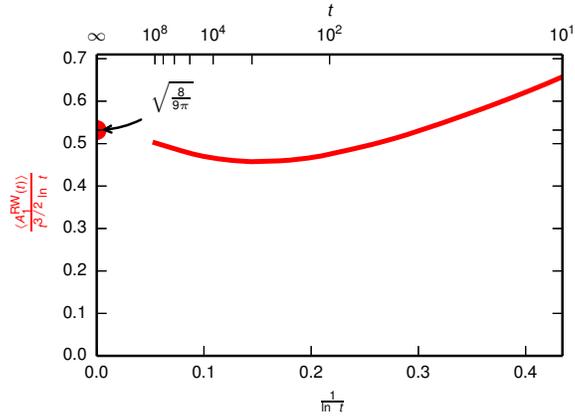}\\
(a)\\
\includegraphics[width=0.5\textwidth]{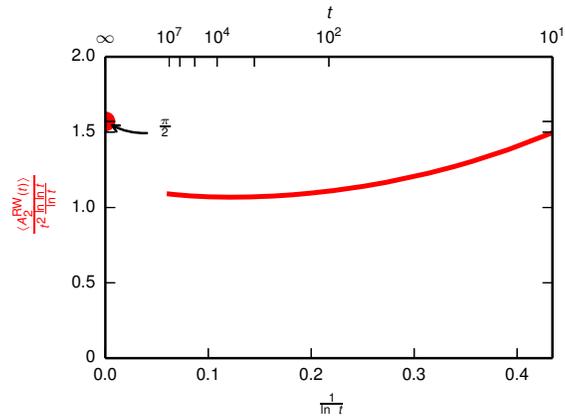}\\
(b)\\
\includegraphics[width=0.5\textwidth]{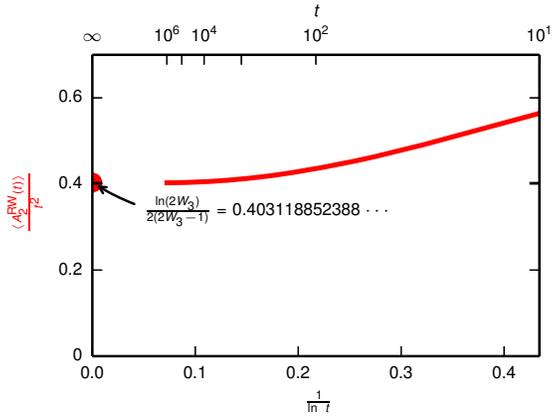}\\
(c)\\
\caption{
Activity (i.e., total number of site visits) for RWs: comparison of results from numerical simulations
with theoretical prediction of asymptotic behaviour.
 (a) One dimension, from 1043 trajectories to $t=10^8$.
 (b) Two dimensions, from 9008 trajectories to $t=10^6$ and 74 to $t=10^7$.
 (c) Three dimensions, from 6689 trajectories to $t=10^5$ and 261 to $t=10^6$.
\label{fig:supplementary-figure-6}
}
\end{center}
\end{figure}

Recall that the total number of visited sites $V(t)$ exhibits
essentially the same behaviour for the SEP and for RWs. In contrast,
the activity is different for these two models. Since RWs hop
independently, the average activity is easily expressed through the
average total number of RWs:
\begin{equation}
\label{act_RW}
\langle A_d(t)\rangle = \int_0^t dt'\,\langle N_d(t')\rangle
\end{equation}
Combining this result with \eqref{Nav:RW} we obtain 
(Figure~\ref{fig:supplementary-figure-6})
\begin{equation}
\label{Aav:RW}
 \langle A_d\rangle \simeq
\begin{cases}
\sqrt{\frac{8}{9\pi}}\,\, t^{3/2}\,\ln t              &d=1\\
\frac{\pi}{2}\,\frac{\ln\ln t}{\ln t}\, t^2           &d=2\\
\frac{\ln(2W_d)}{2(2W_d-1)}\, t^2              &d\geq 3
\end{cases}
\end{equation}

For the SEP, the average activity is
\begin{equation}
\label{act_SSEP}
\langle A_d(t)\rangle = \int_0^t dt'\int d{\bf x}\,[1-\rho({\bf x},t')]\rho({\bf x},t')
\end{equation}
where $\rho({\bf x},t)=\langle n_{\bf x}(t)\rangle$ is the average
density at site ${\bf x}$. Indeed, the activity increases by one
whenever a particle hops into site ${\bf x}$. This happens with rate
$[1-\rho({\bf x},t)]\rho({\bf x},t)$ if the average density varies on
scales large in comparison with the lattice spacing, which is
valid in the large time limit when $d=1$ and $d=2$. Writing the local
activity as $[1-\rho({\bf x},t)]\rho({\bf x},t)$ also tacitly assumes
the validity of the mean-field approximation. In one dimension, for
instance, the exact result for the rate is
$\langle[1-n_x(t)][n_{x-1}(t)+n_{x+1}(t)]/2\rangle$, so even if we can
write $\rho(x,t)\simeq \langle n_{x+1}(t)\rangle\simeq \langle
n_{x-1}(t)\rangle$ we do make the mean-field assumption: $\langle n_x
n_{x+1}(t)\rangle\simeq\langle n_x n_{x-1}(t)\rangle\simeq
\rho^2(x,t)$.

In one dimension the average density is
\begin{equation}
\label{av_den_1d}
\rho(x,t)=\text{erfc}\!\left(\frac{|x|}{\sqrt{2t}}\right)
\end{equation}
where $\text{erfc}(u)\equiv \frac{2}{\sqrt{\pi}}\int_u^\infty dv\,e^{-v^2} = 1 - \text{erf}(u)$ is an error function. 
Substituting \eqref{av_den_1d} into
\begin{equation*}
\langle A_1(t)\rangle = 2\int_0^t dt'\int_0^\infty dx\,[1-\rho(x,t)]\rho(x,t)
\end{equation*}
we obtain 
\begin{equation*}
\langle A_1(t)\rangle = a_1 t^{3/2}, \quad 
a_1 = \frac{4\sqrt{2}}{3} \int_0^\infty dX\,\text{erf}(X)\,\text{erfc}(X)
\end{equation*}
Computing the integral yields $(\sqrt{2}-1)/\sqrt{\pi}$ leading to 
\begin{equation}
\label{a1}
a_1 = \frac{4(2-\sqrt{2})}{3\sqrt{\pi}}=0.440659475\ldots
\end{equation}
which is in excellent agreement with simulations. 

In two dimensions, the average density reads \cite{Krapivsky1992}
\begin{equation}
\label{density_2d}
\rho_{\bf x}(t) = (\ln t)^{-1}\,E_1\!\big(\tfrac{r^2}{t}\big)
\end{equation}
where $r^2\equiv x_1^2+x_2^2$ and $E_1(z)=\int_1^\infty
\frac{du}{u}\,e^{-zu}$ is an exponential integral. Inserting
\eqref{density_2d} into \eqref{act_SSEP} and integrating we get
$\langle A_2\rangle = \pi t^{2}/(2\ln t)$. In three and higher
dimensions, we can still use \eqref{act_SSEP} since the main
contribution to the spatial integral is gathered on large distances,
$r\gg 1$, where the average density varies on the scales large in
comparison with the lattice spacing. The density is small in this
region, so we may simplify the spatial integral: $\int d{\bf
  x}\,[1-\rho({\bf x},t')]\rho({\bf x},t')\simeq \int d{\bf
  x}\,\rho({\bf x},t')=\langle N_d(t')\rangle$ and thereby use the
same Eq.~\eqref{act_RW} as in the case of RWs. Recalling that $\langle
N_d(t')\rangle = (2W_d)^{-1}\,t'$ for the SEP, see \eqref{Nav}, we
get $\langle A_d\rangle \simeq a_d t^2$ with
$a_d=(4W_d)^{-1}$. Collecting these results we arrive at Eq.~\eqref{A:SSEP_num}, repeated here
(Figure~\ref{fig:supplementary-figure-3}):
\begin{equation}
 \langle A_d\rangle \simeq
\begin{cases}
 a_1\, t^{3/2}                              & d=1\\
 \tfrac{\pi}{2\ln t}\, t^2                 & d=2 \\
 (4W_d)^{-1}\, t^2                      & d\geq 3
\end{cases}
\end{equation}

Integrated activity (equivalently the total number of changes of
configurations) has been studied for the SEP on a ring with a fixed
number of particles~\cite{Appert:PhysRevE.78.021122}, where higher
moments and the large deviation function have also been derived. In
our setting, with a source, we have only computed the average. Another
interesting challenge will be to determine the scaling function
$F(M)$.

\section{Summary and outlook}
In summary, we have characterized the asymptotic growth of the average
number of particles injected, the average number of lattice sites they
visit, and the average total number of visit events, for lattice gases
with an infinitely strong point source, in all
dimensions. Additionally, the fluctuations of these quantities are of
interest. For applications to foraging in two dimensions, further work
will be needed to describe the development of the shape of the
visited domain.  Lastly, while for us the model here was motivated by
DNA walkers, it does not account for the memory effect that arises
with catalytic DNA walkers, which modify lattice sites as they walk,
causing hopping rates to change. We hope to return to these
matters in future studies.

\ack{This material is based upon work supported in part by the National Science Foundation under grants CDI-1028238 and CCF-1318833.
We thank Matthew R. Lakin for comments on the draft.}

\appendix

\section{The shape of the domain of visited sites}
One would like to explain 
the apparent relationship $\frac{R_2}{R_1} \sim 1 + \frac{C}{\ln R_1}$.
A fractal structure of the hull
(the external perimeter) of the visited region is also
interesting. For a single RW moving on a 2D substrate, the hull has
fractal dimension $\sfrac{4}{3}$ (this was conjectured by Mandelbrot~\cite{4/3}
and proved by Lawler, Schramm, and Werner~\cite{LSW}). There is
convincing numerical evidence~\cite{Bunde} that the same remains valid
for the hull formed by a fixed number of RWs, and we believe that in
our situation the fractal dimension is also $\sfrac{4}{3}$.  One can ask
topological questions, e.g., how the number of holes scales with time
(in 2D), what is the genus of the surface of the visited domain (in
3D), etc. These questions are very challenging and even for a single
RW little is known~\cite{van_H_a,van_H_b}.

\section{Sub-leading terms}
\label{subs}

The convergence to the leading asymptotic behaviours is often slow, so
one needs an estimate of sub-leading terms, as in Eqs.~\eqref{V3}
and \eqref{V2}, to match theoretical predictions with numerical
results. Here we discuss some other sub-leading corrections, in
particular, we derive \eqref{slow}.

Let us look at the average total number of particles. For the SEP in
one dimension, a neat exact expression \cite{Krapivsky1992} for $\langle
N_1\rangle$ in terms of the modified Bessel functions
\begin{equation}
\label{Nav:exact}
\langle N_1\rangle = e^{-t}\left[I_0(t)+2tI_0(t)+2t I_1(t)\right]
\end{equation}
allows one to extract the leading and sub-leading asymptotic behaviours:
\begin{equation}
\label{Nav_1:asymp}
\frac{\langle N_1\rangle}{\sqrt{8t/\pi}} = 1 + \frac{1}{8t} + \mathcal{O}(t^{-2})
\end{equation}
Thus the convergence is fast, and plotting $t^{-1/2}\langle N_1\rangle$ versus $t^{-1}$ 
one can confirm the amplitude of the leading asymptotic with very high precision
(Figure~\ref{fig:supplementary-figure-1}a).

In three dimensions, one similarly finds
\begin{equation}
\label{N3:sub}
\frac{\langle N_3\rangle}{t}=
(2W_3)^{-1} + \frac{1}{W_3^2}\left(\frac{3}{2\pi}\right)^{3/2} t^{-1/2}+\ldots
\end{equation}
As in one dimension, the sub-leading term plays a small role since the
convergence is still fast. Plotting $t^{-1}\langle N_3\rangle$ versus
$t^{-1/2}$ one can confirm the amplitude $(2W_3)^{-1}$ of the leading
asymptotic with very high precision
(Figure~\ref{fig:supplementary-figure-1}c).

Only in two dimensions is it indeed important to extract the
sub-leading term, yet such a computation is particularly difficult in
two dimensions. Adapting the results from Ref.~\cite{Krapivsky1992} we find that
the Laplace transform of the average total number of particles is
given by
\begin{equation}
\int_0^\infty dt\,e^{-st} \langle N_2(t)\rangle = \frac{1}{s}\left[\frac{1}{sI(s)}-1\right]
\end{equation}
where 
\begin{equation}
\label{Int}
I(s) = \int_0^{2\pi}\frac{dq_1}{2\pi} \int_0^{2\pi} \frac{dq_2}{2\pi} \frac{2}{2s+2-\cos q_1 - \cos q_2}
\end{equation} 
The integral $I(s)$ can be expressed via elliptic integrals
\begin{eqnarray*}
\pi I(s) &=&
\tfrac{2}{1+s}\left[K\left(\tfrac{1}{1+s}\right) - F\left(\sqrt{\tfrac{1+s}{3+2s}},\tfrac{1}{1+s}\right)\right] \\
&+& \tfrac{2}{2+s}\left[K\left(\tfrac{\sqrt{1+s}}{1+s/2}\right) - 
F\left(\tfrac{(1+s/2)\sqrt{1+2s}}{(1+s)^{3/2}},\tfrac{\sqrt{1+s}}{1+s/2}\right)\right]
\end{eqnarray*}
The long-time behaviour of $\langle N_2\rangle$ can be deduced from 
the small $s$ behaviour of its Laplace transform. In the $s\to +0$ limit, one gets
\begin{equation*}
\pi I(s) = \ln(C/s) + \mathcal{O}(s\ln(1/s)), \quad C=16\sqrt{3}-8
\end{equation*}
Therefore
\begin{equation}
\int_0^\infty dt\,e^{-st} \langle N_2(t)\rangle =\frac{\pi}{s^2\ln(C/s)}+\ldots
\end{equation}
in the $s\to +0$ limit, from which
\begin{equation}
\langle N_2(t)\rangle =\frac{\pi t}{\ln(tCe^\gamma)}
\end{equation}
proving the result announced in \eqref{slow}.

\bibliographystyle{ieeetr}
\bibliography{paper}

\end{document}